\documentstyle[preprint,eqsecnum,aps]{revtex}
\begin{document}
\draft
\tighten
\preprint{SU-GP-98/6-1}
\preprint {gr-qc/9806123}
\title{Spacetime Embedding Diagrams for Black Holes\thanks{e-mail:marolf@suhep.phy.syr.edu}}
\author{Donald Marolf}
\address{Physics Department, Syracuse University, Syracuse, New York 13244}
\date{June, 1998}
\maketitle

\begin{abstract}
We show that the 1+1 dimensional reduction (i.e., the radial plane)
of the Kruskal black hole
can be embedded in 2+1 Minkowski spacetime and discuss how 
features of this spacetime can be seen from the embedding diagram.
The purpose of this work is educational:
The associated embedding diagrams may be useful for
explaining aspects of black holes to students who are familiar with
special relativity, but not general relativity.  

\end{abstract}
\pacs{COMMENT: To be submitted to the American Journal of Physics.  
Experts will wish only to skim
appendix A and to look at the pictures.}

\input{epsf.tex}
\vfil
\eject

\baselineskip = 16pt
\section{Introduction}

The diagram below appears in many popular treatments of black holes
in books, magazines, and science museums.

\centerline{\epsfbox{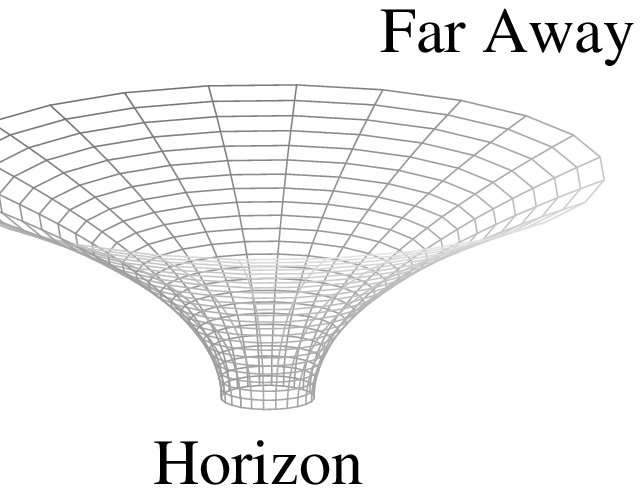}}

\centerline{Fig. 1:  The exterior $t=const$ equatorial plane of a Schwarzschild
Black Hole.}

\smallskip

\noindent It accurately represents
a certain aspect of a Schwarzschild black hole, namely the intrinsic
geometry of a two dimensional surface in the space around the black hole.  
The particular
surface involved is the equatorial plane at some 
(Killing) time\cite{MTW837}.  That is to say that
the analogous diagram for the earth would describe the geometry of
the plane indicated in fig. 2 below.
Fig. 1 shows how the equatorial plane of a black hole
would be curved if, instead of lying in the black hole spacetime,
it were part of familiar flat Euclidean three space.  
Mathematically, this picture is said to represent an 
{\it embedding} of this plane into
three dimensional Euclidean space.

\centerline{
\epsfbox{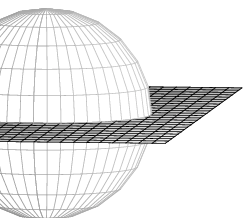}}

\centerline{Fig. 2: The equatorial plane of the Earth.}

\smallskip

Diagrams such as fig. 1 can be useful for explaining certain
geometrical features, such as the fact that circles drawn around the 
equator of the black hole barely change in size as they are pushed
inward or outward near the horizon.  However, since they
refer only to space at a single instant of 
time, such diagrams do not describe the
most important parts of black hole physics having to do with the
space{\it time} structure of the geometry.

In addition, such pictures
can be misleading for the uninitiated.  For example, since the bottom of the
funnel in fig. 1 is the black hole horizon, students may be tempted
to believe that the horizon represents a real boundary of the spacetime.
Another problem is that many students believe they can see the gravitational
attraction of the black hole in fig. 1 by visualizing the path
of a ball `tossed onto the funnel.'  Technically speaking however, 
there is no direct connection between fig. 1 and the attraction of a black
hole:  there are spacetimes with the same exact spatial geometry as a black
hole in which freely falling objects maintain a constant 
position (with respect to static worldlines), 
and even such spacetimes
in which objects fall in the wrong direction\footnote{For example, in
the metric $ds^2 = -dt^2 + {{dr^2}\over{(1-2M/r)}} + r^2 d\theta^2
+ r^2 \sin^2 \theta d\varphi^2$, worldlines with constant $r, \theta, \varphi$
are freely falling.  In the spacetime
$ds^2 = -{{dt^2}\over{(1-2M/r)}}+ {{dr^2}\over{(1-2M/r)}} + r^2 d\theta^2
+ r^2 \sin^2 \theta d\varphi^2$, free fallers that begin at rest with respect to
this coordinate system fall toward large $r$.}!

The purpose of the present work is to fill in this gap by creating diagrams
which do show the spacetime structure of the black hole and which can therefore
be used to explain this structure to students.  These
new diagrams will make it clear that the horizon is in fact
much like any other part of the spacetime.  They will also allow one to
see the real gravitational `attraction' of the black hole.

Specifically, the aim is
to provide a means to take students familiar only with special
relativity and {\it show} them a number of features of black holes.
This work itself is presented at a slightly higher level and is intended
for readers with some familiarity with the basic concepts of General
Relativity and black holes, though the technical material is relegated
to the appendices.  A reading of, for example, one of \cite{Geroch,MV,Kauffman}
should provide adequate preparation, and even the reader
familiar only with special relativity should be able to gain some understanding
from sections \ref{diag} and \ref{SSsec}.

To make our new kind of diagram, we will again choose a two
dimensional surface 
and embed this surface in a three dimensional
space.  However, since we wish to show the space{\it time} aspects
of the geometry, our new surface will include both a spacelike direction
and a timelike direction.  That is, some pairs of events on our surface will
be separated by a timelike interval while other pairs will be separated
by a spacelike interval.  We 
therefore need to draw the surface inside three dimensional (2+1)
Minkowski space as opposed to Euclidean three space.  This means that
our diagrams are most useful for students with a strong grasp of special
relativity.

Technically, 
we will work with what is known as the Kruskal spacetime
(also known as the analytically
extended Schwarzschild spacetime) \cite{Wald}.  
This describes an `eternal' black hole
which does not form by the collapse of matter but instead
has been there forever.
In particular, it describes a spherically symmetric such black hole with
no electric charge or angular momentum.

The surface we consider here is the radial, or $rt$, plane given by
$\theta=const$, $\varphi=const$ in a spherical coordinate system. 
Due to the spherical symmetry,
it contains the worldline of any observer with zero angular momentum, whether
falling freely or accelerating in a rocketship. This is also
true for (fictitious) observers traveling faster than light.
The radial plane is often (e.g. \cite{dimred,CGHS}) 
called the dimensional reduction of the 3+1
black hole to 1+1 dimensions as it is effectively a 1+1 version of a black
hole\footnote{Note, however, that it is slightly different from
the 1+1 dilaton black holes of, for example, \cite{CGHS,GS}.}.

A light review of the Kruskal spacetime is presented in section \ref{Krusk}
below.  The interested reader should consult \cite{Wald,MTW826}
for a more thorough discussion.
The embedding diagram is shown in section \ref{diag}, and we
discuss there how it may be used to illustrate a number of general
features of the black hole geometry.  We save those features associated
with the horizon or with the Schwarzschild coordinate system for section
\ref{SSsec}.  Because of its more technical
nature, the derivation of the equations describing the embedding
has been placed in appendix \ref{rt}.  Section \ref{disc} contains
a short discussion of the results and some extensions.
Finally, appendix B comments briefly on the embedding of
other surfaces: the radial plane of spacetimes describing
star-like objects and the analogue of fig. 1
{\it inside} the black hole.

\section{A Brief Review of the Kruskal Spacetime}
\label{Krusk}

As stated above, the Kruskal spacetime describes an eternal black hole
which exists forever.  Such black holes have much in common with
the usual astrophysical sort that form from the collapse of matter, but
they are mathematically simpler.  On the other hand, they also have
certain odd features which their astrophysical cousins do not share.
For example, the Kruskal spacetime contains not just one, but
two separate `asymptotic regions' on opposite sides of the black hole,
connected by an Einstein-Rosen bridge\cite{MTW837}.
Thus, the old kind of embedding diagram (the
analogue of fig. 1)
for this case involves two copies of fig. 1 glued together at the horizon
as shown in fig. 3. 
The left and right parts are often described as two separate
Universes connected by the `wormhole' (Einstein-Rosen bridge) in the middle.
In contrast,
an astrophysical black hole has only one exterior region.  As a result,
only a part of the Kruskal spacetime (containing, say, the right, but not
the left side of fig. 3) would be relevant to a
discussion of such black holes.  More will be said about this below.

\centerline{\epsfbox{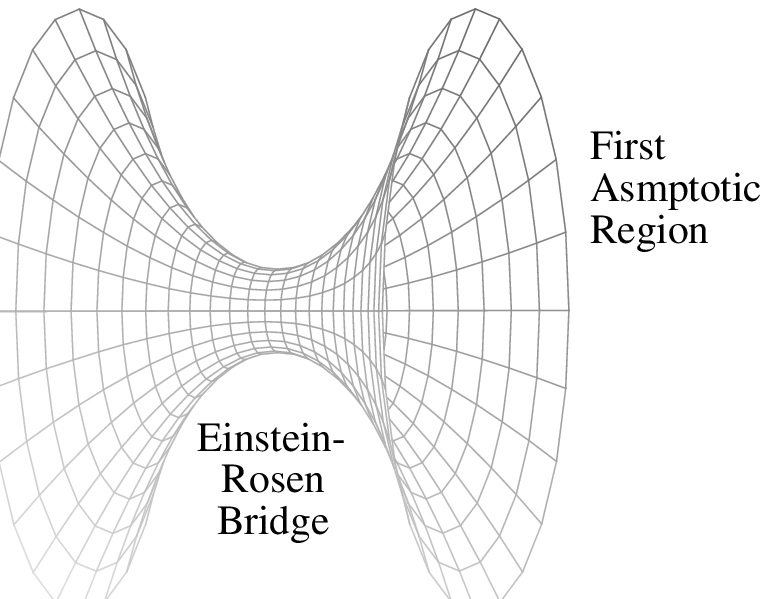}}

\centerline{Fig. 3:  The $t=const$ 
equatorial plane of a Kruskal black hole.}

In fig. 3, we have rotated both the left and right parts of the diagram with
respect to fig. 1.  This serves to illustrate the fact that 
the orientation of these diagrams
in space carries no information.  It will also allow the reader
to more easily relate fig. 3 to the diagrams that appear below, which are
of a different sort for which the orientation in space does carry information.

We remind the reader that the global
structure of the Kruskal spacetime is summarized \cite{MTW834} by the Penrose
diagram in fig. 4.  This diagram shows only the time and radial directions
and does not include the angular directions around the black hole.
In particular, time runs up and down and space runs left to right.
This means that, 
figs. 3 and 4 have only one direction in common: the spatial direction 
which runs more or less right to left on both diagrams.

Recall that a Penrose diagram does not accurately portray distances and times, 
but it {\it is} drawn so that light rays always travel along lines at
$45$ degrees to the vertical, no matter how the spacetime is curved.
This is a very useful property that in fact all of our remaining diagrams 
will share.  Note that the solid lines in fig. 4 all correspond to the paths
of light rays.  

\centerline{\epsfbox{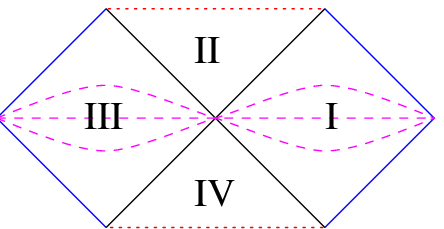}}
\centerline{Fig. 4: The Penrose diagram for a Kruskal black hole.}

\smallskip

Regions I and III of fig. 4 are two disconnected `exterior' regions.
Each dashed line shows the location of a surface like the one
shown in fig. 3.  The Kruskal spacetime contains infinitely many such surfaces,
all exactly alike and each lying half in region I and half in region III
with the `throat' just at the point where the two regions touch.
As mentioned above, only one of these external regions
(say, I) has an analogue in astrophysical black holes.

The other regions are referred to as the interior of the black hole:
region II is the `future interior' and region IV is the `past interior.'
Since light rays travel at $45$ degrees to the vertical, objects
traveling at less than the speed of light can neither enter region IV nor
leave region II.  As a result, the light rays that form the boundaries
of these regions are the horizons of the black hole.
There are also two singularities, one in the past in region IV
and one in the future in region II, indicated by the dotted lines
on the diagram.  Note that the singularities are {\it spacelike} lines.

The past interior (region IV) is another part of our spacetime with no
analogue in an astrophysical black hole.  The diagram below
should clarify the relationship between such black holes and the
Kruskal spacetime:

\smallskip

\centerline{\epsfbox{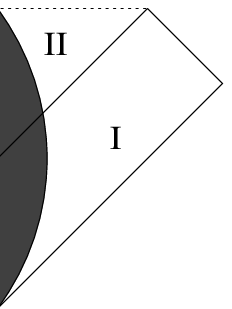}}
\centerline{Fig. 5:  The Penrose diagram for a black hole that forms
from stellar collapse.}

\noindent
The unshaded parts of regions I and II are essentially the
same as regions I and II in the Kruskal black hole.  The shaded
region is the star that collapses to form the black hole.  Inside
the star, the spacetime is slightly different from Kruskal, though except for
removing regions III and IV it is still qualitatively very similar.

We now turn to a description of the
Schwarzschild coordinate system, which is a useful way to label events
in the Kruskal spacetime.  Briefly, this system 
consists of four functions, $r$, 
$t$, $\theta$, and $\varphi$.  The last two of these are the usual
coordinates on spheres `around' the black hole, with $\theta$ being a
longitude-like
coordinate taking values in $[0,\pi]$ and $\varphi$ a latitude coordinate
taking values in $[0,2\pi]$.  On the other hand, $r$ and $t$ are
coordinates that are constant on such spheres and whose values pick
out which particular sphere one is referring to.  The function $r$ tells
how {\it big} the particular sphere is; the area of the sphere is
$4\pi r^2$.  An important subtlety is that $r$
does not directly give the distance away from {\it anything}.  In addition, 
the Kruskal spacetime does not have a `center' at $r=0$ in the usual
sense.  The last function, $t$, is related to a symmetry of the Kruskal
spacetime.
Its most important property is that, in the exterior (regions I and III),
surfaces of constant $t$ are surfaces of simultaneity for observers
who do not move with respect to the black hole.  

The Schwarzschild coordinate system cannot really be used on the entire
Kruskal spacetime as it breaks down at the horizon.  However, it can
be used {\it separately} in the interiors of regions I, II, III, and IV.
When this is done, one of the unusual features of this coordinate system is
that $t$ is only a timelike coordinate in the exterior (regions I and III).
In the interior (regions II and IV), it is spacelike.  Similarly, the
coordinate $r$ is spacelike in the exterior (regions I and III) where it
takes values $r > 2MG/c^2$.  Here,
$M$ denotes the mass of the black hole, $G$ is Newton's universal
constant of gravitation, and $c$ is the speed of light.  
In the interior (regions II and IV), $r$ is timelike and
takes values $r < 2MG/c^2$.  
On the horizon itself, $r$ is a lightlike coordinate and takes the
constant value $r=2MG/c^2$.  Lines of constant $r$ and $t$ are
shown in fig. 6 below in all four regions:

\smallskip

\centerline{\epsfbox{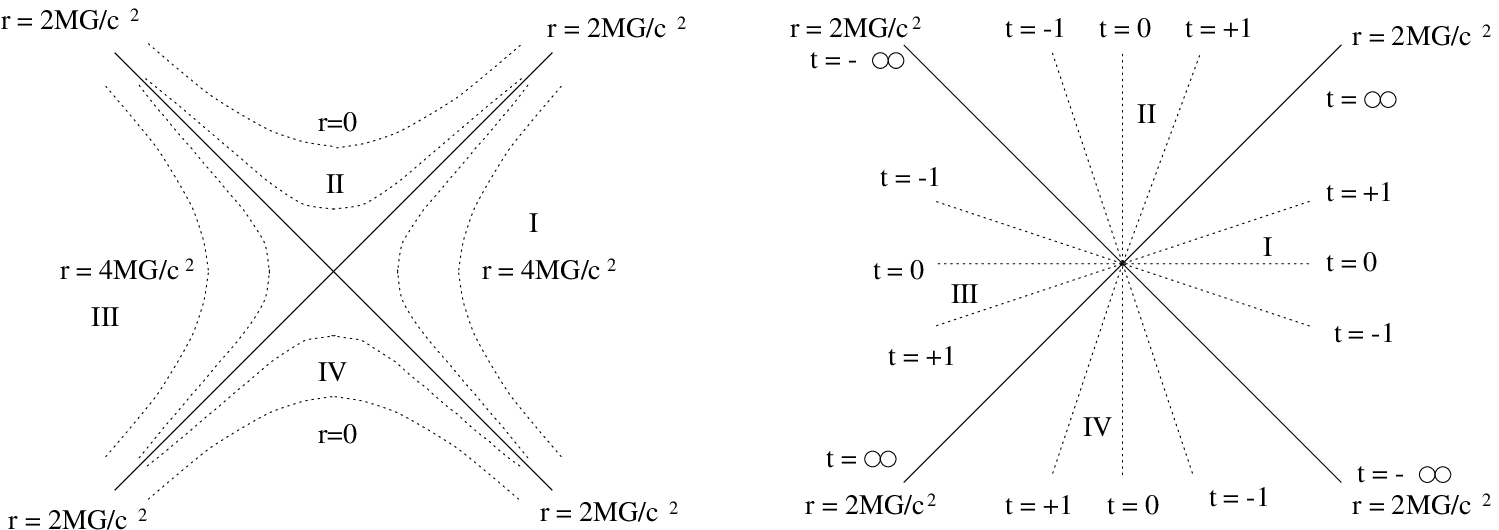}}
\centerline{Fig. 6: The Schwarzschild coordinate system.}

\noindent  Again, these diagrams 
are drawn so that, despite the curvature of the spacetime, light rays 
always travel at 45 degrees to the 
vertical.  Note that the fact that the $t=0$ line runs through
the point where the horizons cross does not make it a `special' time in
any way since all of the other $t=const$ lines run through through this
point as well.  In fact, any transformation $t \rightarrow t + constant$
is a symmetry of the black hole.  In the diagrams of fig. 6, this
symmetry looks like a boost transformation around the origin.

As a more technical reminder, we mention that the actual geometry 
of the Kruskal spacetime is encoded in the metric which, in
Schwarzschild coordinates, takes the form\cite{MTW820}

\begin{equation}
\label{SS}
ds^2 = - c^2 \left(1 - 2MG/rc^2 \right) dt^2 + {{dr^2} \over {1 - 2MG/rc^2}} +
r^2 d\theta^2 + r^2 \sin^2 \theta d\varphi^2.
\end{equation}
This form
holds in the interior of every region (I, II, III, and IV) but cannot
be used directly at the horizons (where $r = 2MG/c^2$).

\smallskip

\section{The embedding diagram}
\label{diag}

The diagram below shows the embedding derived in appendix \ref{rt}.  
It illustrates what the radial (i.e., 
$\theta=const, \varphi = const$) plane of a Kruskal black hole would
look like if it were a surface in 2+1 Minkowski space.  As with
the Penrose diagram of fig. 4, it describes only the time and radial
directions, and not the angular directions.

\epsfbox{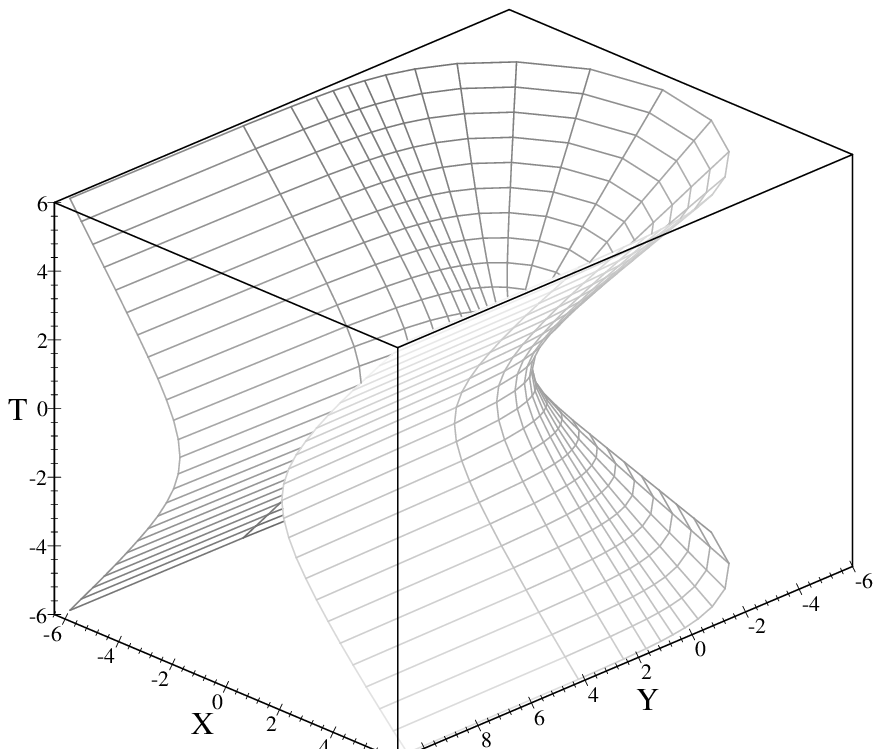}

\centerline{Fig. 7:  The $\theta=\varphi=const$ embedding diagram
for a Kruskal black hole,}
\centerline{in units where $c=1$ and $2MG/c^2 =1$.}

Here, the vertical ($T$) direction is timelike and
the horizontal ($X$ and $Y$) directions are spacelike.
Thus, time again runs up and down while space runs across the diagram.
Units have
been chosen in which the speed of light ($c$) is one, so that light
rays travel at $45$ degrees to the $T$ axis.  In addition, the size of the
black hole has been fixed by using units in which the `Schwarzschild
radius' of the horizon ($2MG/c^2$) has been set to one.  
The lines in fig. 7 serve only
to guide the eye and do not represent any
particular structure of the black hole solution.  

Note that the
diagram is a smooth surface, with little to distinguish one point
from another.  In particular, it is not immediately clear which
points lie on the horizon $r=2MG/c^2$.  This is an excellent way to show
students that the horizon is not essentially different from any other
part of the spacetime.

As the reader may wish to view and manipulate this surface\footnote{After
generating a three dimensional plot, Maple allows one to grab the diagram
with the mouse and turn it around by hand.} for
herself, we mention that this picture was drawn with a five line Maple
code.  The basic idea of the code is to introduce a (not quite smooth)
coordinate $s$ on the embedded surface so that $s,T$ each range
over the real line and so that they change rapidly where the surface
is flat, but slowly where the surface is highly curved.  The following
code is compatible with both releases 4 and 5 of MapleV4:

\medskip

$>$ restart;

$>$ X := (s,T) -$>$ tanh(5*s/2)*sqrt(s $\hat{}$ 2 + T $\hat{}$ 2);

$>$ sigma :=  (s,T) -$>$ (X(s,T)) $\hat{}$ 2 - T $\hat{}$ 2 ;

$>$ Y := (s,T) -$>$ evalf( Int( sqrt( ( (1-q) $\hat{} $(-4) - 1) /q), 
q=0..sigma(s,T)/4),2);

$>$ plot3d([Y(s,T),X(s,T),T],s = -1.9 ..1.9, T = -5..5, scaling=constrained,

axes=box, labels=[Y,X,T], orientation=[45,65], axesfont=[TIMES,ROMAN,11],

labelfont=[TIMES, ROMAN, 18]);

\noindent
The integration variable $q$ corresponds to $\sigma/16M^2$ from
appendix \ref{rt}. 

\medskip

Let us now examine the basic features of our diagram.
Note that it has two long `flanges' (marked below in fig. 8)
which project out in the positive
$Y$ direction.

\centerline{\epsfbox{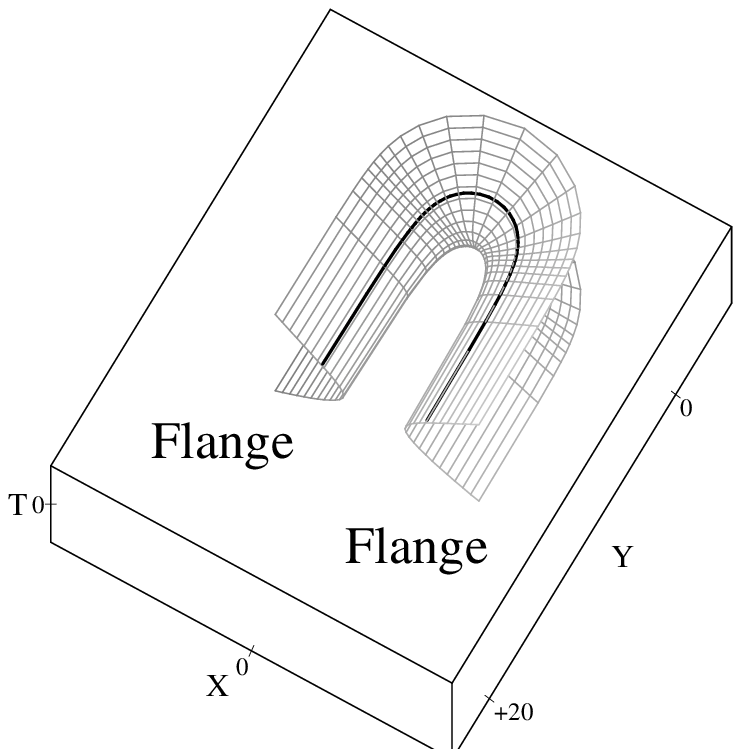}}
\centerline{Fig. 8:  The asymptotic regions.}

\noindent
Moving along the flanges in the +Y direction (say, 
along the heavy line in fig. 8 above) allows one to move a large proper
distance in a spacelike direction.  Thus, these flanges must correspond
to the asymptotic regions far away from the black hole.  This is also
clear from equation (\ref{Y}) in appendix A.  We may, for example, take the
right flange to represent region I of the Penrose diagram (fig. 4) and
the left flange to represent region III.

Now, far from the black hole (that is, far out on these
flanges), spacetime should be flat.  We can see that this is so from
the above pictures.  If we move far enough out along a given flange, 
the black line in fig. 8 becomes straight.  Thus, the flanges are
curved only in the $XT$ plane and not in the $YT$ or $XY$ planes.
A surface that is curved only in
one direction 
(when embedded
in a flat three space) has zero intrinsic curvature \cite{spivak},
e.g., a cylinder is flat. 
In this way, we can see that the intrinsic curvature of
our slice vanishes as we move away from the black hole.

One of the most important uses of this sort of diagram is that it does 
allow one to see the gravitational
`attraction' of the black hole.  Specifically, it
allows one to see the worldlines followed by freely falling observers, 
and to see that they curve toward the middle of the diagram, away from
the asymptotic regions.  The point is that, in General Relativity, 
freely falling observers follow `geodesics,' the straightest possible
lines on a curved surface.  Human beings are in fact quite good at
visualizing such lines, just by pretending that they are `walking'
up the surface shown in the embedding diagram\footnote{An observer must
walk more or less
`up' the diagram, as we assume that the motion is forward in time and
at less than light speed.} (fig. 7).  For example, visualizing a person
walking around a sphere shows that geodesics on a sphere are just
the great circles\footnote{The present
case is slightly different as we deal with a curved surface
in Minkowski space and not in Euclidean space.  This will
lead the human eye to make some errors, but they are typically small.} 

Thus, let us consider a freely falling observer moving up one of the flanges.
Because our surface bends {\it away} from the ends of the flanges and
{\it toward} the center, our observer will follow this curve and also
move closer to the center of the surface.  A computer generated geodesic
of this type is shown in fig. 9 below.

\centerline{\hbox{\epsfbox{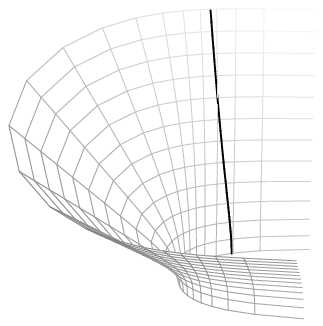}}}
\centerline{Fig. 9:  The worldline of an observer who falls freely from rest
at $r=3MG/c^2$,}
\centerline{starting at T=0.}

Between the two asymptotic regions, the surface is curved.  Note that
the central portion of the surface rapidly approaches a light cone:

\centerline{\epsfbox{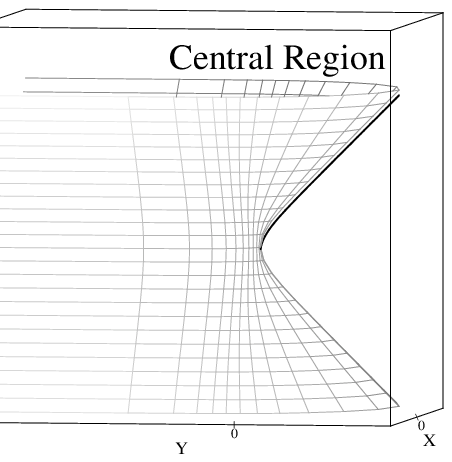}}
\centerline{Fig. 10: The central portion rapidly approaches a light cone.}

\noindent
It turns out that this is what
accounts for the fact that we see no direct
visual sign of the black hole singularity.
Equation (\ref{rs}) from appendix A shows that the singularity $r=0$ is 
in fact located at $T = \pm \infty$ in the distant future and past.  
The point here is that our surface approaches
a light cone (along which no proper time would pass) so quickly that, if we
follow a line up the center of this diagram (such as the black line at
the right edge of fig. 10 above), it reaches $T=\pm \infty$
after only a finite proper time.  Thus, although the black hole singularity
lies at the boundary of the spacetime, an observer will reach it in
a finite proper time.  A nice feature of the diagram is that
it shows that the singularity does not occur at any {\it place} inside
the black hole.  Instead, it is better thought of as occurring at some
{\it time}, which in this case happens to be $T = \pm \infty$. 

Now, a tricky point of the diagram is that, despite the fact that
$r=0$ is a curvature singularity, the surface
appears to become flat as we follow it toward the singularity at
$T = \pm \infty$.  This is
a result of the fact that the surface is drawn from a fixed reference
frame in the 2+1 Minkowski space while observers traveling in the surface
are (for large $T$) moving at nearly the speed of light with respect
to that frame.  For example, the worldline in fig. 11 is moving much
faster (relative to the frame in which the diagram is drawn) at event 
E5 than at event E1. In fact, each successive event drawn corresponds
to an increase of the boost parameter ($\tanh^{-1}(v/c)$) by $0.5$.

\centerline{\epsfbox{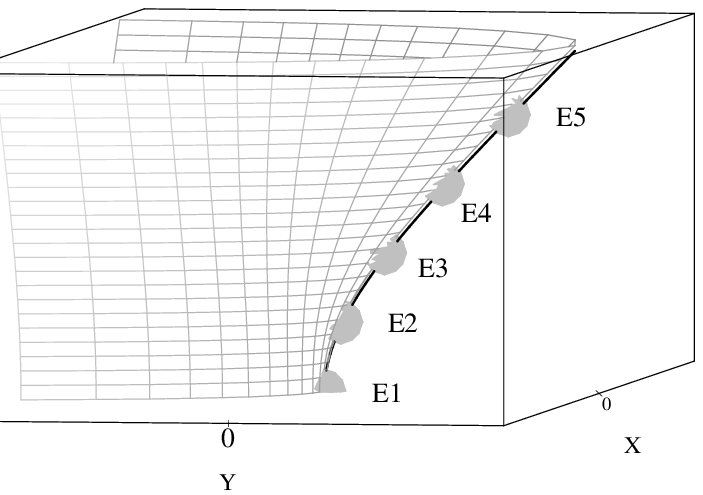}}
\centerline{Fig. 11: Five events along a freely falling  worldline.}

The familiar effects of time 
dilation and length contraction
have the effect of `flattening out' the diagram.  The fact
that a finite bit of proper time is expanded to reach all the way to
$T = \pm \infty$ means that, for large $T$, we are given such a
close up view of the surface that the curvature is not visible.  To 
actually {\it see}
the curvature at some event with our eyes, we must redraw the picture in
a set of reference frames in which an observer at that event is at rest.  Thus, 
we must consider a series of more and more
highly boosted reference frames.  Below, we have used a series of boosts
in the $YT$ plane to redraw the embedding diagram in the reference frame
of the wordline from fig. 11 at each of the events E1, E2, E3, E4, and E5.  

\hbox {\epsfbox{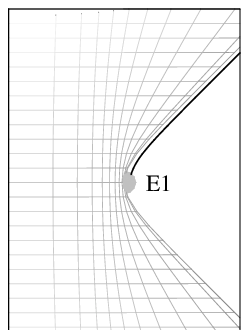} \epsfbox{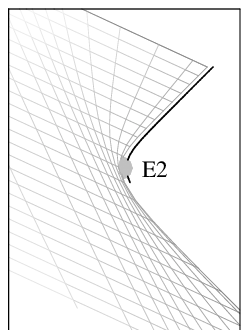} 
\epsfbox{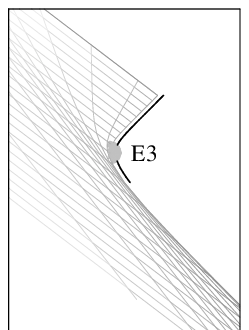}
\epsfbox{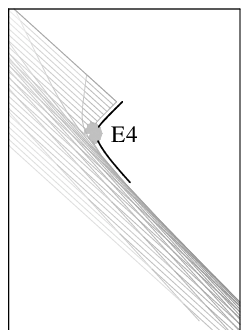} \epsfbox{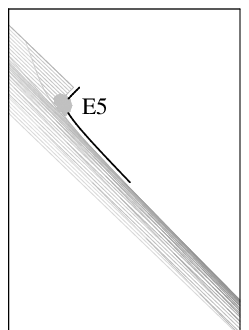}}

\vskip.2cm 

\centerline{Fig. 12: The embedding diagram viewed from the side and
boosted into the} 
\centerline{reference frame of the indicated worldline at
E1, E2, E3, E4, and E5.}

The singularity is now evident in the fact that, as we increase
the boost and so examine points closer and closer to its location, 
the `corner' in the diagram becomes sharper and sharper.  This
shows that the intrinsic curvature of the surface becomes larger and larger as
$T$ increases and $r$ goes to zero.  Physically, we can see that geodesics
which are close to each other before the corner diverge much more
quickly after passing E5 than after passing E1.  The diagrams in fig. 13 below
each show the same two geodesics, but from the reference frames of
E1, E2, E3, E4, and E5 respectively.

\hbox {\epsfbox{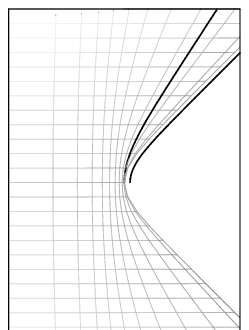} \epsfbox{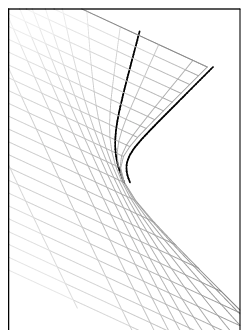} 
\epsfbox{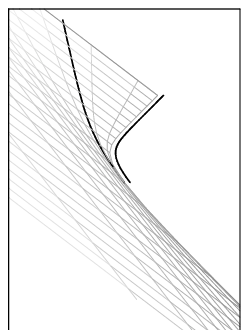}
\epsfbox{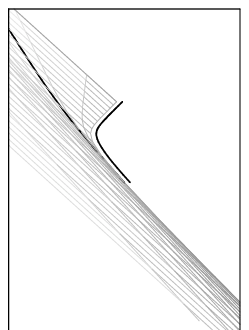} \epsfbox{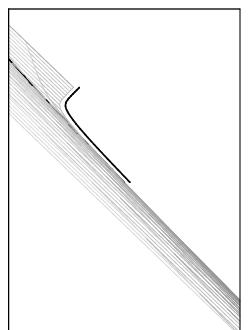}}

\vskip.2cm 

\centerline{Fig. 13: The intrinsic divergence of two nearby worldlines
near different events.}

\section{The Horizon and Schwarzschild Coordinates}
\label{SSsec}

It is instructive to discover the horizons of the black hole
by directly examining the embedding diagram without making reference
to the equations in appendix \ref{rt}.  Recall that the future
horizons are the edges of the region (II) from which an observer
cannot escape without traveling faster than light.  Similarly, the past
horizons are the boundaries of the region (IV) that an observer (starting far
away) cannot enter without traveling faster than light.  

Note that the two light rays given by $Y=0$, $X = \pm T$ (shown in
fig. 14) lie completely in our surface.
The reader will immediately see that these light
rays do not move along the flanges at all (since they stay at $Y=0$)
and thus neither of these light rays actually move away from the black hole.
Instead, the light rays are trapped near the black hole forever.
It is also clear that these light rays divide our surface into
four regions (I, II, III, and IV) much as in figs. 4 and 6.  Thus, 
an observer in region II of our spacetime cannot cross one of these light
rays without traveling faster than the speed of light.  Observers
in this region are trapped inside the black hole.

\centerline{\hbox{  \epsfbox{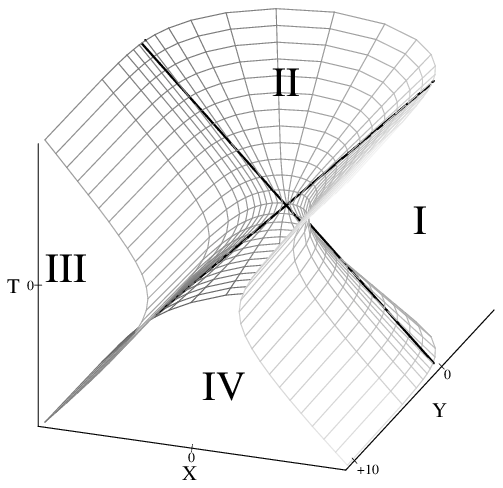} \epsfbox{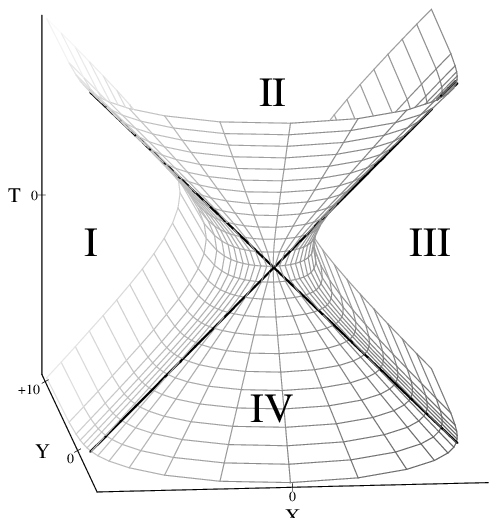}}}
\centerline{Fig. 14:  The lines $Y=0$, $X = \pm T$ divide the diagram
into four regions.}

On the other hand, any light ray fired outward from region I or III of our 
surface {\it will} eventually reach large values of $Y$ and thus escape
from the black hole.  Thus, the light rays at $Y=0$, $X = \pm T$ (for
$T >0$) are the future horizons of the black hole.  Similarly, these same
light rays for $T < 0$ form the past horizons of the black hole.
This may also be seen from equations (\ref{Y}) and (\ref{rs}) in
appendix A.
  
We now turn again to the
Schwarzschild coordinates $r$ and $t$ which were briefly reviewed in
section \ref{Krusk}.
Insight into this coordinate system
can be gained by drawing these
coordinates directly on our embedding diagram:

\hbox{\epsfbox{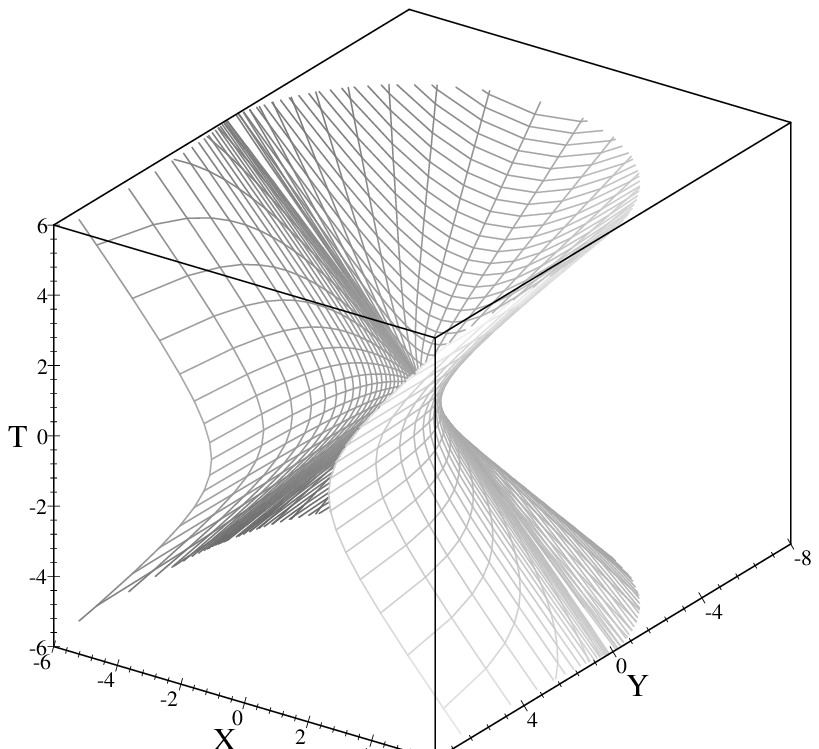} \epsfbox{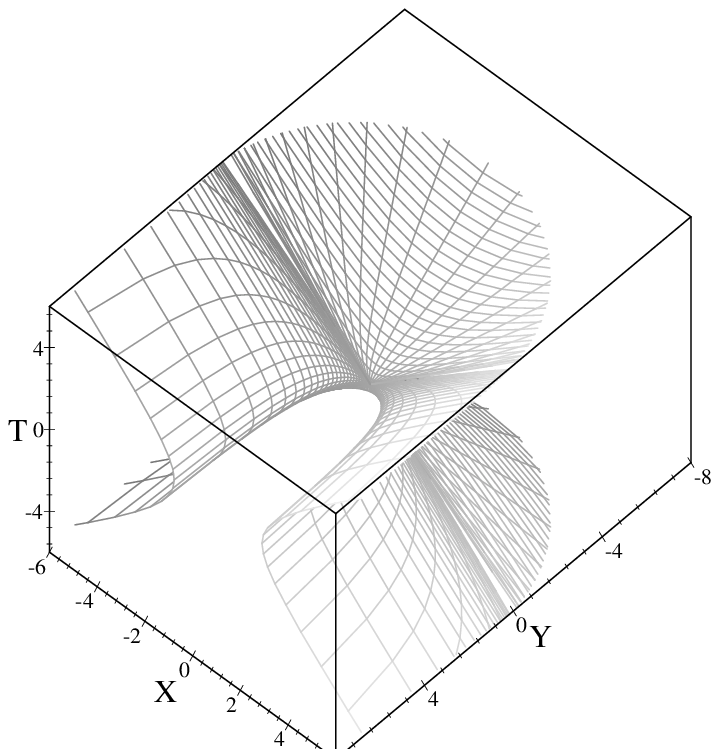}}

\vskip .2cm
\centerline{Fig. 15: Two views of the embedding diagram 
with Schwarzschild coordinate lines.}

\smallskip

\noindent
This has also been done
with a short Maple code.  In this case, it is easiest to first have Maple
draw each region (I,II,III,IV)
separately, as we can then use $r,t$ as intrinsic coordinates in
the surface and Maple's plotting algorithm will draw the lines of
constant $r,t$ for us.  The four pieces can then be combined into a single
diagram.  Thus, region I can be drawn (with MapleV4 release 4 or 5) using
the code:

$>$ restart;

$>$ X1 := (s,f) -$>$ s*cosh(f); T1 := (s,f) -$>$ s*sinh(f);

$>$ Y1 := (s,f) -$>$  evalf(Int( sqrt(( (1-q) $\hat{}$ (-4)
-1)/q),q=0..(s $\hat{}$ 2)/4),2);

$>$ bh1:=plot3d([Y1(s,f),X1(s,f),T1(s,f)], s=0..1.84, 
f=-3..3, scaling=constrained,

axes=box, labels=[Y,X,T],axesfont=[TIMES,ROMAN,11], 

labelfont=[TIMES,ROMAN, 18]); bh1;

\noindent
For those who have read appendix A, we mention that  
f is the hyperbolic angle $\phi$ and s is proportional to
the proper distance $\rho$, which is a function of $r$.
In this way, lines of constant s or f are
constant $r$ and $t$ 
lines respectively\footnote{Of course, the constant $r$ lines
drawn in this way are not equally spaced contours of $r$.  Rather, they
are equally spaced contours of $s$.  If one wishes, this is easily corrected
by replacing $s$ with $\sqrt{\left| 1 - {{2M} \over r} \right|}$.  However, 
this makes it more difficult to see the geometry of the surface, as
$r$ changes slowly in some regions where the surface is tightly curved.}. 
Again, the integration variable $q$ corresponds to $\sigma/16M^2$
and the diagram is shown in units where $2MG/c^2 =1$; thus, $s^2 =
\sigma/4M^2$.

Region II is drawn with the code:

$>$ restart;

$>$ X2 := (s,f) -$>$ s*sinh(f); T2 := (s,f) -$>$ s*cosh(f);

$>$ Y2 := (s,f) -$>$ evalf(Int( sqrt(( (1-q) $\hat{}$ (-4)
-1)/q),q=0..(-s $\hat{}$ 2)/4),2);

$>$ bh2:=plot3d([Y2(s,f),X2(s,f),T2(s,f)],s=0..6,
f=-3..3, scaling=constrained,

axes=box, labels=[Y,X,T],axesfont=[TIMES,ROMAN,11], 

labelfont=[TIMES,ROMAN, 18]); bh2;

\noindent
Regions III and IV can be drawn in the same way, or obtained by an
appropriate rotation of regions I and II.

Recall from fig. 6 that the horizon lies at $t = \pm \infty$ in Schwarzschild
coordinates.
Since
Maple is not be able to draw the surface all the way to $t = \pm \infty$,
there is a slight gap in the diagram at the horizon which serves
to illustrate its position. 
The horizon is also clearly marked (even in the region where
no gap is visible) as a large ``X'' due to the fact that the
lines of constant $t$ pile up near $t=\pm \infty$.

As discussed in section \ref{Krusk}, the
lines of constant $t$ are spacelike in regions I and III
(where $t$ is a timelike coordinate) and are timelike in regions II and IV
(where $t$ is a spacelike coordinate).  Thus, the lines running across
the flanges (in regions I and III) are constant $t$ as are the lines
running up and down the central light cone (in regions II and IV).
The black lines in fig. 16 below show the location of a few
constant $t$ lines in each region.  As seen previously in fig. 6, 
all lines of constant $t$ intersect at a common point, in our case
at $X=Y=T=0$.

\centerline{\epsfbox{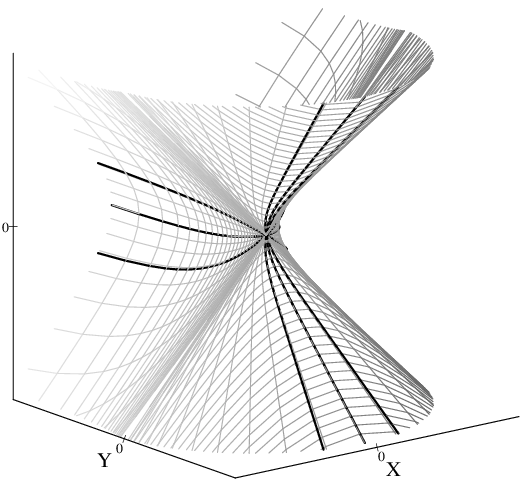}}
\centerline{Fig. 16:  Lines of constant $t$ in each
region.}

On the other hand, lines of constant $r$ run up the flanges in regions
I and III and across the light cones in regions II and IV, as shown by
the black lines in fig. 17.  A useful property of these lines to know
is that a given line of constant $r$ lies entirely in a plane of constant
$Y$, as is clear from the second diagram in fig. 17.  
In particular, as was already noted, 
the horizons are located at $r=2MG/c^2$ and lie in the $Y=0$ plane.
The $Y> 0$ region has $r > 2MG/c^2$ and corresponds to 
outside the black hole, while the $Y<0$ region has $r < 2MG/c^2$ 
and corresponds to the inside.  As was already stated, $r=0$
is the singularity and is located inside the horizons at $T = \pm \infty$.
The reader will notice a strong similarity between the first diagram in
fig. 17 and the symmetry
transformation of fig. 8.  This is because the symmetry does not 
$r$ and therefore only slides events back and forth along lines of constant
$r$.

\centerline{\hbox{\epsfbox{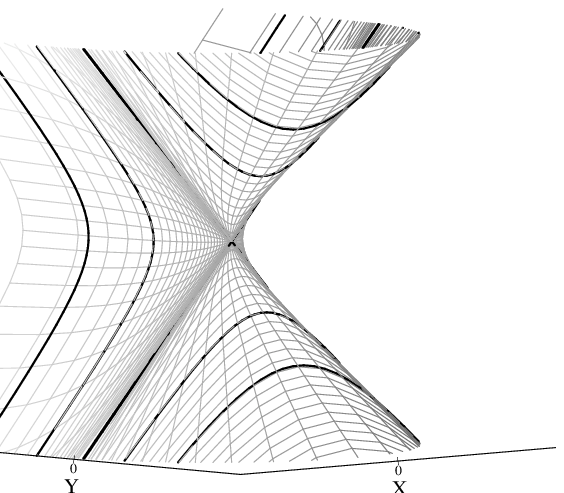} \epsfbox{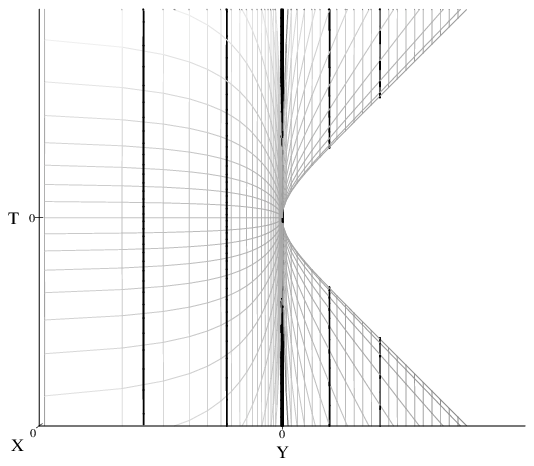}}}
\centerline{Fig. 17: The lines $r = 8MG/c^2$, $r=4MG/c^2$, $r=2MG/c^2$, $r=MG/c^2$,}
\centerline{ and
$r=MG/2c^2$.}

Such diagrams
are useful for seeing the large acceleration (tightly
curved worldlines) of observers
who remain at constant $r$ just outside the horizon.  Here, we should perhaps
pause to remind the reader which aspects of our diagram represent
`physical' effects and which do not.  The important point here is
that the ambient Minkowski space is {\it only} a mechanism for visualizing
the surface and, by itself, carries no physical information.  In particular, 
the way that a worldline bends relative to the ambient Minkowski space is not
important by itself, so long as the entire surface is bending in the same
way.  What {\it is} important, and what does correspond to the 
proper acceleration
of a worldline in the black hole spacetime is that way that the worldline
bends {\it within} our surface.  

Now, 
the constant $r$ lines in fig. 17 all bend only in the $XT$ plane; they
remain at a constant value of $Y$.  This means that their proper
acceleration vector (in the ambient Minkowski space at, say, $T=0$) points
in the $+X$ direction.  Now, close to the middle of our diagram, 
the $+X$ direction is more or less tangent to our surface.   
Thus, the proper acceleration has a large
component tangent to our surface, and these worldlines have a large
proper acceleration {\it in} the black hole spacetime.  
This corresponds to the fact that a spaceship
which is close to the black hole must fire its rockets with a large
thrust in order not to fall in.  In contrast, 
far away from the black hole the $+X$ direction is almost completely
orthogonal to our surface at $T=0$.  
Thus, the proper acceleration of such worldlines {\it within}
our surface is virtually zero, and spaceships far enough away from the 
black hole need only the slightest thrust to avoid falling in.

Another useful feature of the Schwarzschild coordinates is that they
illustrate the gravitational time dilation (redshift) that occurs near a
black hole.  
Recall from section \ref{Krusk} that the $t=const$ lines are lines
of simultaneity for the family of (accelerated) observers who follow
worldlines of constant $r$).  Now, note that (as shown in fig. 18 below)
the spacing between the $t=const$
lines outside the black hole varies with $r$, so that static observers at
different values of $r$ have clocks that accumulate proper time
at different rates:

\smallskip

\centerline{
\epsfbox{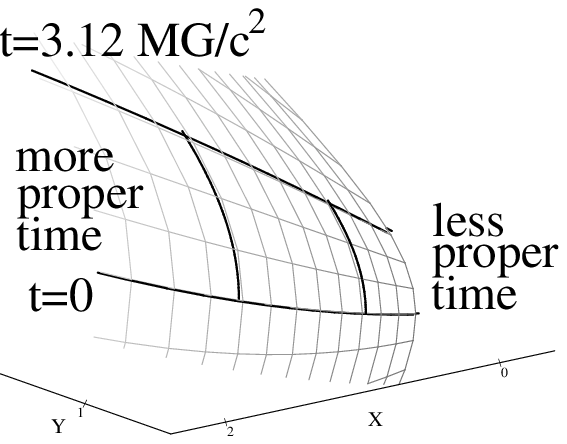}}
\centerline{Fig. 18:  Static observers experiencing more and less 
proper time}
\centerline{between two constant $t$ lines.}

As a final feature, we point out that the infinite stretching of objects
due to the gravitational tidal forces in the radial direction near the
singularity can also be seen from fig. 15.  The point here is
that a line of constant $t$ (in the interior) is a timelike
geodesic.  Thus, it is the worldline of some freely falling observer.
Consider two such lines, one at $t=t_0$ and one at $t= t_0 + \delta$, 
such as the ones shown in fig. 19:

\centerline{
\epsfbox{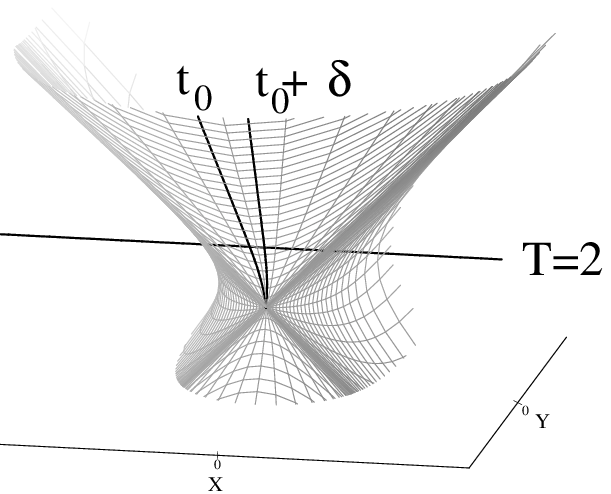}}
\centerline{Fig. 19: Worldlines following $t=t_0$ and $t=t_0 + \delta$.}

\noindent
For small $\delta$, these observers are very close to each other and
are nearly at rest relative to each other at, say, $T=2$.  On the other hand, 
we can see that, if we wait until $T = +\infty$ (when they hit the
singularity), they will be infinitely far apart (say, as measured along
a line of constant $r$).  But $T = + \infty$ is only a short {\it proper}
time in the future!  Thus, the two geodesics will separate infinitely far
in a finite proper time.  It follows that their relative acceleration (or, 
equivalently, the gravitational tidal `force') diverges as $T \rightarrow 
\infty$.

\section{Discussion}
\label{disc}

We have constructed an embedding of the radial plane of the Kruskal black
hole in 2+1 Minkowski space and used it to illustrate certain features
of black hole spacetimes.  Such diagrams may be of use in describing
black holes to students familiar with special, but not general, relativity.
In particular, such diagrams make it clear that the horizon is a smooth
subsurface of the spacetime and show the gravitational attraction of the
black hole.  

Unfortunately, not all 
two dimensional surfaces with one spacelike and one timelike direction
can be embedded in 2+1 Minkowski space.
This is much the same as trying to embed two dimensional Riemannian
spaces in three dimensional Euclidean space (and it is well known that, 
for example, the two dimensional surface of constant negative curvature
cannot be so embedded).  Thus, a completely general spacetime cannot
be treated in the same way.  However, appendix \ref{star} shows that it
is in fact possible to embed the radial plane of any spacetime which 
describes a `normal' static and spherically symmetric star-like object
(without horizons or regions of infinite density).  
As an example, for a `star' made of a thin spherical shell of mass
$M$ located at $r = 4MG/c^2$, the associated embedding diagram is:

\centerline{
\epsfbox{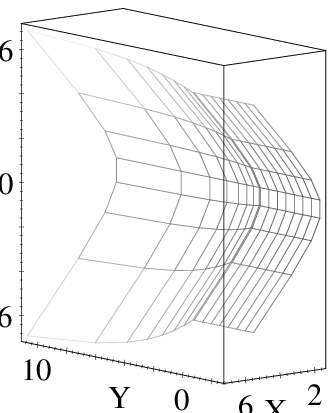}}
\centerline{Fig. 20: The embedding diagram for a static thin shell of mass $M$
at $r =  4MG/c^2$,}
\centerline{shown in units with $c=1$ and $2MG/c^2 = 1$.}

\noindent
The shell is located at the `crease' and lines of constant Schwarzschild
radius $r$ and time $t$ have been shown.  Such diagrams may serve
to give students an intuitive picture of the way that matter curves
spacetime in General Relativity.  In particular, it may be of interest
to study such diagrams for stars with different equations of state 
to get a feeling for the different gravitational effects of pressure
and density.

More general embeddings of black hole spacetimes with horizons should be
possible as well.
In particular, a study of the embedding of
the radial plane of the Reissner-N\"ordstrom black hole
is in progress.

\acknowledgements 
I would like to thank Alan Middleton for his assistance with Maple and Peter
Saulson for helpful comments and advice.  The author would like to
express a special thanks to
Charles Misner and Dieter Brill for finding (and fixing!) an incompatibility
between a previous version of the Maple code and MapleV4 release 5.
This work  was supported in part by NSF grant
PHY-9722362 and funds provided by Syracuse
University.

\appendix

\section{Embedding the Radial Plane}
\label{rt}

In this appendix, we derive the equations that describe the embedding
shown in fig. 7.  Recall that we consider 
the radial plane given by setting the angular coordinates
$\theta, \varphi$ to (arbitrary) fixed values and letting $r,t$ range
over all of their values.  The motion of any
observer which maintains zero angular momentum must take place in such
a plane whether the observer falls through the horizon or remains outside.
Since $d\theta =0 = d\varphi$ on this plane, the metric on our surface is given
by the first two terms of (\ref{SS}):
\begin{equation}
ds^2 = -\left( 1 - {{2M} \over r} \right) dt^2 + {{dr^2} \over {1 -
{{2M} \over r}}}.
\end{equation}
In this equation, and in the rest of the appendix, we use units in
which Newton's constant ($G$) and the speed of light ($c$) are
set equal to one.

Let us begin with region I.
To construct the embedding, we will make use of the symmetries of the
spacetime.  This region has a time translation symmetry given by 
$t \rightarrow t + \Delta t$.  However, we see from the metric
(\ref{SS}) that, under such a symmetry transformation, not all points
on the surface are displaced through an equal amount of proper time.
Instead, a point at coordinate $r$ is displaced through a proper time
$ \Delta t  \sqrt{1 - 2M/r}$.  This is the familiar gravitational redshift 
near the horizon of a black hole.

Thus, in order to embed region I of our surface,
we will need to find a timelike symmetry of 2+1 Minkowski space which also
displaces different points through different amounts of proper time.
Let us endow our 2+1 Minkowski space with Cartesian coordinates $X,Y,T$ in the
usual way.  One such symmetry is a boost of the Minkowski space, say, in
the $X,T$ plane.  Of course, this symmetry is not timelike everywhere
on Minkowski space, but induces a division of this spacetime into
four regions in a manner analogous to figs. 4 and 6.  The boost is
timelike in regions I and III and spacelike in regions II and IV.
Since the time translation on the Kruskal spacetime has the same property, 
this boost is a good candidate for the analogue of a time translation in
our embedding space.  We expect each of the regions I,II,III,IV of the radial
Kruskal plane to embed in the corresponding region of Minkowski space.

Having realized this, it is useful to introduce coordinates on Minkowski
space that are adapted to the boost symmetry.  In region I ($X > |T|$), let
\begin{eqnarray}
\rho &=& \sqrt{X^2 - T^2}, \cr
\phi &=& \tanh^{-1}(T/X).
\end{eqnarray}
These are just the usual proper distance and hyperbolic angle in region I
of the $X,T$ plane.   In these coordinates, the metric on our Minkowski
space is
\begin{equation}
ds^2 = - \rho^2 d \phi^2 + d\rho^2 + dY^2
\end{equation}
and a boost corresponds to $\phi \rightarrow \phi + \Delta$. Thus, 
$\phi$ should be proportional to the $t$ of our embedded plane.
We will choose to set $\phi = {t \over {4M}}$ as this choice will lead
to a smooth embedding.

It follows that a boost that sends $\phi$ to $\phi +
\Delta$ corresponds to a time translation that
sends $t$ to $t + 4M \Delta$.  This moves a point at $\rho$
in the Minkowski space by $\rho \Delta$ and a point at $r$ in the 
Kruskal spacetime by $(1 - 2M/r)^{1/2} 4M \Delta$.  Therefore, 
we must set $\rho = 4M \sqrt{1 - 2M/r}$.    To complete our
construction of the embedding, we need only give $Y$ as a function of
$r$ and $t$.  By symmetry, it can in fact depend only on $r$.  The
function $Y(r)$ is determined by the requirement that the metrics
agree on a $t = 4M \phi = const$ slice:

\begin{equation}
\label{curv}
{{dr^2} \over {1 - 2M/r}} = ds^2 = d\rho^2 + dY^2.
\end{equation}
Solving this equation yields

\begin{equation}
\label{Y}
Y(r) = \int_{2M}^r \sqrt{\left( 1 + {{2M} \over r} + {{4M^2} \over {r^2}}
+ {{8M^3} \over {r^3}} \right)} dr.
\end{equation}
Here, we have chosen the horizon $r=2M$ to be located at $Y=0$.
It may now be checked that the induced metric on the surface 
$\phi = t/4M$, $\rho = 4M \sqrt{1 - 2M/r}$, $Y = Y(r)$ in region I is given
by the first two terms of (\ref{SS}).  Note that, for any $r$, $\rho^2 < 
16M^2$.

This completes our embedding of region I of the radial Kruskal plane
in region I of the 2+1 Minkowski space.  The embedding of region III
is exactly the same, except that we work in region III of the
Minkowski space so that $X < 0$.
In regions II and IV, we introduce $\rho$ as the proper time 
from the origin in the $T,X$ plane and $\phi$ as the associated hyperbolic
angle $\tanh^{-1}(X/T)$.  The discussion differs from that of regions I and III
by an occasional sign at the intermediate
steps, so that, for example, $\rho = 4M \sqrt{{{2M}\over r} -1 }$. However, 
in the end we arrive at exactly the same expression (\ref{Y})
for $Y(r)$.  This feature depends on the precise choice of the
proportionality factor
between $t$ and $\phi$ and justifies the choice ($4M$) made above.

Thus, we have smoothly embedded the four regions I,II,III,IV of the
Kruskal spacetime in 2+1 Minkowski space.  These four pieces in fact
join together to form a single smooth surface.  To verify this, we need only
show that, on our surface, 
one of the embedding coordinates (say, $Y$) is a smooth
function of the other two ($X,T$).  We have written $Y$ as a smooth
function of $r$, 
so we must now study $r$ as a function of $X$ and $T$.  We have
written $r$ in a slightly different way in each region but, if we introduce
$\sigma = X^2 - T^2$, we see that 
\begin{equation}
\label{rs}
r = {{2M} \over {1 - \sigma/16M^2}}
\end{equation}
everywhere on the surface.  Since $X^2 - T^2 < 16M^2$ for every point on
our surface, $r$ is a smooth function of $X$ and $T$.
It follows that $Y$ is also a smooth function of $X$ and $T$ and that
our surface is smooth.

Maple codes which can be used to plot this surface with and without
Schwarzschild coordinate lines were given in sections \ref{diag} and
\ref{SSsec} respectively.  The numerical integral in these codes
is just equation (\ref{Y}), though it takes a somewhat different form as
it has been written using $q=\sigma/16M^2$ instead of $r$ as
the integration variable.

\section{Embedding other surfaces}

In this appendix, we consider the embedding of other surfaces in 2+1
Minkowski space.  Section \ref{star} shows that the radial
plane of `normal' static spherically symmetric star-like spacetimes can be
so embedded.  Thus, diagrams like fig. 7 can
be generated for a large class of interesting spacetimes.  
On the other hand, section
\ref{rtheta} returns to the Kruskal spacetime, but considers a different
totally geodesic surface, in this case one
associated with observers who orbit the black hole but always remain
inside the horizon.  The pedagogical use of this latter embedding diagram
is limited, but it does illustrate the crushing gravitational tidal
`force' of the black hole in the angular directions. 

\subsection{The radial planes of star-like objects}
\label{star}

The metric for a general
static spherically symmetric spacetime (without horizons) can be written
in the form \cite{MTW604}
\begin{equation}
ds^2 = - e^{2\Phi} dt^2 + {{dr^2} \over {1 - 2m(r)/r}} + r^2 d \Omega^2,
\end{equation}
where $m(r) = \int_0^r 4 \pi r^2 \rho dr$ is the
total mass-energy contained inside the radius $r$ and the potential
$\Phi$ satisfies
\begin{equation}
{{d\Phi}  \over {dr}} = {{ m + 4 \pi r^3 p} \over {r(r-2m)}}.
\end{equation}
Here, $\rho$ is the energy density, $p$ is the pressure, and 
we have set $c = G = 1$.  For a
static `normal' star without horizons
(or incipient horizons), we expect $p$, $\rho$, and $(1-2m/r)^{-1}$ to
be bounded.
Under these conditions, and assuming asymptotic flatness, 
the quantity $e^\Phi \left( {{d\Phi} \over {dr}} \right) \sqrt{1-2m/r}$
is bounded, say by $Q$.  We may then embed the radial plane of any such 
metric in region I of 2+1 Minkowski space by setting
$\phi = Q t$ and following the same procedure used in section II.
The function $Y(r)$ is determined by the equation
\begin{equation}
\label{dYdr}
dY/dr = \sqrt{{1 \over {1-2m/r}} - e^{2\Phi} \left({{d \Phi} \over {dr}}
\right)^2 Q^{-2} },
\end{equation}
and, given the bound above, the square root is real.  The resulting
embedding diagram may be drawn by evaluating $\Phi$, setting
$\rho = e^ \Phi/Q$ (and $s = \rho/4M$), and adjusting the definition of
Y1 in the Maple code used for black hole region I in section \ref{SSsec} to
solve (\ref{dYdr}), taking $Y1=0$ at some arbitrary radius $r_0$.

\subsection{The $r-\varphi$ plane of the Kruskal black hole}

\label{rtheta}

Another totally geodesic surface in the Kruskal spacetime is the analogue of
fig. 1 inside the black hole.  This is the surface given
by $t=const$ and $\theta = const$ for $r < 2MG/c^2$, 
and its geometry is independent of
either $t$ or $\theta$.  All geodesics with zero momentum in the $t, \theta$
directions which start in this surface must remain there.
Since we are considering the surface {\it inside}
the black hole, this is again a timelike surface and we will wish to
embed it in 2+1 Minkowski space.  This can be done
simply by taking $r,\varphi$ to be the usual radius and
polar angle in the $XY$ plane
and setting $r = 2M (1 - ({T \over {4M}})^2)$.  
In fig. 21 below
we have used a slightly different plotting scheme than in the main text
for esthetic reasons.

\vbox{
\centerline{\epsfbox{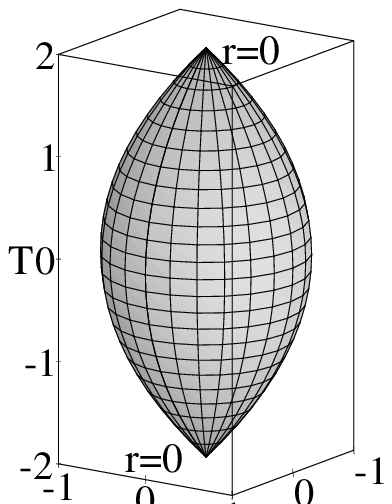}}

\vskip .2cm
\centerline{Fig. 21: The $t=const$, $\theta = const$ surface inside the
Kruskal black hole,}
\centerline{in units where $c=1$ and $2MG/c^2=1$.}}

It is perhaps best to think of each circle above as representing a sphere
of symmetry in the Kruskal geometry.
Here we can see these spheres expanding through region III from zero size
at the past singularity, reaching a maximum at $T=0$ (the horizon, and
in particular the bifurcation sphere -- where regions I,II,III, and IV
all meet), and then contracting again though region I back down to zero size.
In this picture, the singularities 
appear at finite
times ${T\over {2M}} = \pm 2$ in the diagram and are clearly seen as corners
at the top and bottom.
However, we know that, in this surface also, the scalar curvature
should diverge as we approach the singularity.  This is not evident
from the diagram and in fact (except for the corner at the singularity
itself), the diagram looks rather flat in this region.  What is
happening is that the lines of constant $\theta$ in this surface are, 
once again moving at nearly the speed of light as seen from our
reference frame ($dr/dT \rightarrow 1$ as $T \rightarrow 4M$).  As
a result, the diverging scalar curvature has once again been
flattened out by an even more rapidly diverging boost.

On the other hand, the physical effects of the diverging curvature {\it are}
quite clear.  Consider, for example, a steel ring placed `around'
the black hole at $T=0$.  As time passes, if no part of this ring has
any momentum in the directions transverse to the circle, the ring must
simply move up the diagram.  Because the entire diagram contracts
to a point, the ring likewise must contract (i.e., be crushed) no
matter how great a stress the ring can support\footnote{In `real life,'
of course, slight asymmetries would first cause the ring to buckle into 
the directions not shown on this diagram.  In any case, the ring would be
destroyed, no matter how strong it is.}!  This is one effect of the
infinite crushing tidal gravitational `force' in the direction
around the black hole near the singularity.

\end{document}